\documentclass[a4paper,11pt]{article}
\usepackage{pos}
\usepackage{float}
\usepackage{lineno}
\usepackage{graphicx}

\usepackage{pos}
\newcommand*{\mkblue}[1]{{\color{black}{#1}}}
\newcommand*{\mkred}[1]{{\color{black}{#1}}}

\title{Progress of the GRANDProto300 Project}

\author*[a]{PengXiong Ma}
\author[a,b]{Yi Zhang}
\author[a,b]{Xing Xu}
\author[a,b]{Bohao Duan}
\author[a]{Shen Wang}
\author[a,b]{Kewen Zhang}
\author[c]{Pengfei Zhang}
\author[c]{Xin Xu}
\onbehalf{for the GRAND Collaboration{\normalsize \\ \normalfont(a complete list of authors can be found at the end of the proceedings)}\\}

\affiliation[a]{Key Laboratory of Dark Matter and Space Astronomy, Purple Mountain Observatory, Chinese Academy of Sciences\\
  No. 10 Yuanhua Road, Nanjing, China}

\affiliation[b]{School of Astronomy and Space Science, University of Science Technology of China\\
No. 96 Jinzhai Road, Hefei , China}
\affiliation[c]{School of Electronic Engineering,Xidian University\\
No. 2 South Taibai Road, Xi'an, China}

\emailAdd{mapx@pmo.ac.cn}

\abstract{GRANDProto300 (hereafter referred to as GP300) is a pioneering prototype array of the GRAND experiment. It consists of 300 radio antennas and will cover an area of \(200\, \text{km}^2\) in a radio-quiet region of western China. Serving as a test bench for the GRAND experiment, GRANDProto300 aims to achieve autonomous radio detection and reconstruction of highly inclined air showers. It is designed to detect ultra-high-energy cosmic rays in the energy range of \(10^{16.5} - 10^{18}\, \text{eV}\) at a rate comparable to that of the Pierre Auger Observatory. Over the past two years, significant improvements have been made to both the hardware and firmware of GP300. Currently, 65 antenna units have been deployed at the site by June 2025. We present the current status of detector commissioning, including updates on hardware, calibration results such as GPS timing and antenna positioning. Additionally, we discuss the solar radio bursts associated with solar flares, the galactic radio emissions detected, and preliminary cosmic ray surveys.}

\ConferenceLogo{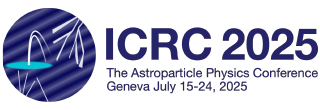}

\FullConference{39th International Cosmic Ray Conference (ICRC2025)\\
 15–24 July 2025\\
Geneva, Switzerland\\}

\begin{document}
\maketitle

\section{Introduction}

Radio detection for cosmic rays above 100 ‌PeV‌ has been emerging in recent ‌decades‌, which is regarded as a crucial way to measure the ‌properties‌ of cosmic rays ‌with accuracy comparable to‌ the developed fluorescence technique‌,‌ the latter ‌of which‌ is designed to measure the electromagnetic component of extended air showers. Moreover, radio detection has ‌the‌ advantage of duty cycle and flexibility for large-scale deployment, and is crucial for detection of very low flux density at ‌the‌ highest energy band.
GRAND is proposed for detecting ultra-high energy cosmic rays (UHECRs) and neutrinos\cite{GRAND:2018iaj,olivier:2025icrc}. A 300-antenna array in western China‌,‌ GRANDProto 300 (GP300)‌,‌ is the pioneering prototype array for GRAND concept. There have been‌ 65 detection units (DUs) deployed at XiaoDuShan site in Dunhuang‌,‌ ‌a‌ city in ‌the‌ north-western ‌part‌ of China‌,‌ in Gansu ‌Province‌. 
In this contribution, we will present the progress in ‌the‌ last two years for the deployment of 65 DUs and the whole system. \mkred{More contributions for this conference from GRAND and the relevant are seen\cite{grandauger:2025icrc,arsen:2025icrc,pablo:2025icrc,aur:2025icrc,Jolan:2025icrc,sei:2025icrc,marion:2025tdh,lukas:2025tdh}. More detailed hardware status for GRAND and progress in radio frequency chain can be found \cite{GRAND:2024atu,GRANDhardware:2025atu}. }

\section{Site surveys}

The first candidate ‌site‌ in China was ‌Lenghu‌ in Qinghai‌, where a few prototypes of DU were deployed in 2019.  At the end of 2019 the situation for GRAND ‌changed‌ during the pandemic. The site survey was ‌restarted‌ in 2021 ‌\mkred{coordinated}‌ by ‌the‌ Purple ‌Mountain‌ Observatory. After ‌reassessing‌ the ‌Lenghu‌ site and ‌conducting‌ new surveys across other possible locations in Qinghai and Gansu‌,‌ both the local government of Dunhuang and ‌the‌ survey group decided to choose ‌XiaoDushan‌ as the experiment site for GP300‌, \mkred{a 2.5-hour drive from the city of Dunhuang, in the Gansu Province, China.}
This site ‌has been‌ confirmed ‌to have‌ a long-term radio quiet background from 50 to 200 MHz \mkred{in last two years}.

\section{Detection unit (DU)}
\begin{figure*}[]
\centering
\includegraphics[width=4.8cm]{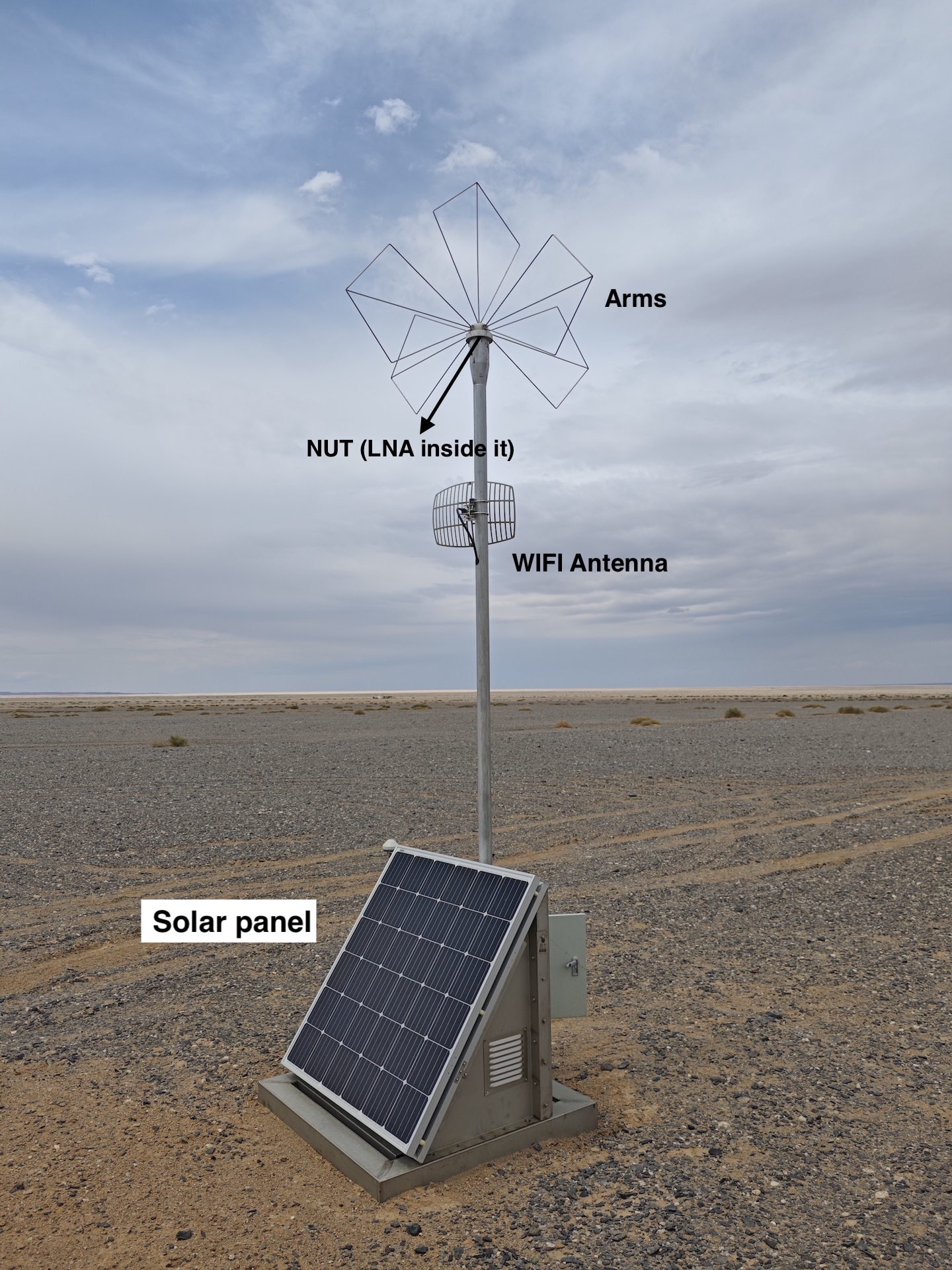}
\includegraphics[width=4.8cm]{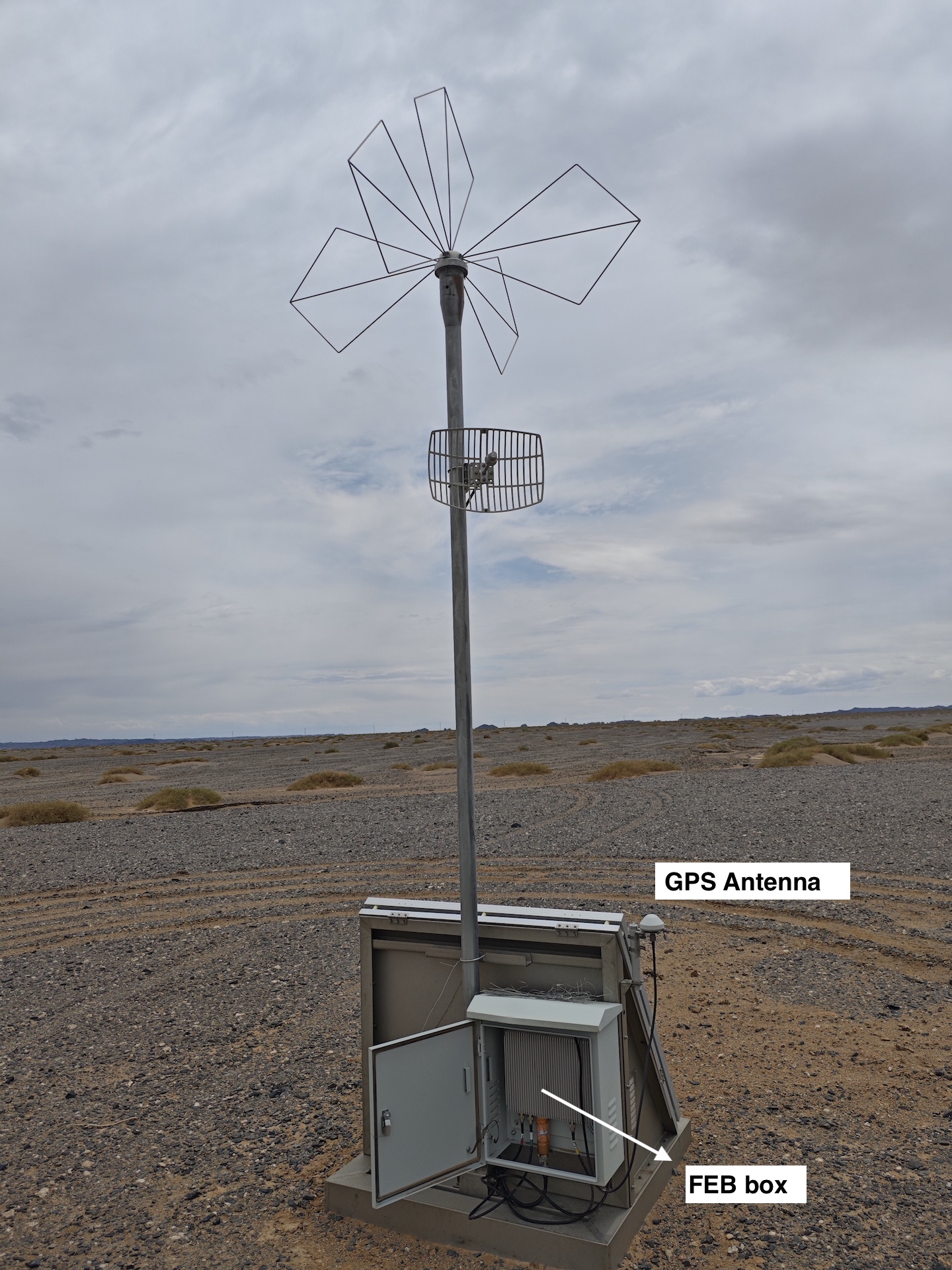}
\caption{Face view (left) and back view (right) of one detection unit of GP300 deployed at the XiaoDuShan site in China.}
\label{fig:du}
\end{figure*}

One DU of GP300 is ‌shown‌ in Fig.\ref{fig:du}‌, which‌ consists of a ‌triangle-shaped‌ base for ‌storing‌ the battery inside ‌it, a solar panel for charging on the ‌south-facing plane surface‌, a 3.5‌-meter-tail‌ pole, and five arms ‌aligned‌ with ‌the‌ north-south, west-east, and vertical directions \mkred{defined as X, Y, and Z polarization, respectively.} There is a nut at the top of ‌the‌ pole‌,‌ with ‌a‌ Low Noise Amplifier (LNA) and matching network mounted‌ inside ‌it‌. A mesh antenna at ‌the midpoint‌ of ‌the‌ pole is used for communication between ‌the‌ DU and ‌the‌ central station. The timing for each DU is ‌managed‌ with a GPS antenna mounted on the base. 
\mkred{The Front End Board (FEB) is mounted behind the triangular shaped box, with updated heat dissipation methods. It accepts three-channel signals from the LNA, each of the signals is amplified by a Variable Gain Amplifier (VGA) and shaped through a 30-200 MHz bandpass-filter circuit before digitized by a 14-bit 500MSPS analog-to-digital-converter (ADC). The digitized signals are moved into a System on Chip (SoC) (Xilinx, Zynq Ultrascale+, XCZU7CG), which performs tasks such as event triggering, event building, communication, and data buffering. After triggering, the firmware running inside the PL part of the SoC builds the event data package which is composed of the data from all three channels, the timing information from the GPS chip and other slow control parameters. Afterwards, the data package is stored into the local DDR memory by the PS part of the SoC.} 

\mkred{The communications between DUs and central station have been implemented by using a rocket (installed at central station) and the WIFI mesh antenna mounted at the pole as shown in Fig. \ref{fig:du} working at 5GHz band in wireless. Current total throughput bandwidth is roughly 150Mbps, which enables us to take data at tens of Hz for trigger event.}

\section{Deployment Progress of GRANDProto300}
The campaigns for GP300 separated into different stages benefit the hardware and software development for GRAND experiment, and also the understanding of the detection method. 

\subsection{GP13 from February 2023 to October 2024}
The ‌first‌ deployment in 2023 for GP13 demonstrated the beginning of GP300 construction‌, whose‌ layout ‌consisted of‌ three hexagons and one central station for data acquisition seen black points in Fig. \ref{fig:deployment}. \mkred{The monitoring data obtained} from the 13-antenna ‌array‌ allowed us to control the temperature ‌overheating‌ inside ‌the‌ FEB box, which is ‌a Faraday‌ box ‌that contains‌ the FEB \mkred{and is} the heart of ‌the‌ DU. Additional heat ‌dissipation‌ function was introduced‌, and‌ it ‌has been‌ proved to work well‌, ensuring‌ the long-term ‌operation‌ on a Gobi site. The solar panel works with a charge controller for battery charging‌. 
Thanks to the commissioning ‌run‌ of GP13, we identified different sources of noise from the whole DU and suppressed or ‌shielded‌ them with ‌appropriate‌ actions‌.‌ More ‌details‌ will be summarized in a journal paper.

GP13 was not dedicated to ‌detecting‌ cosmic rays due to ‌its‌ very small area and ‌hardware/software \mkred{being in an early prototype phase}. ‌However,‌ data ‌acquisition‌ became more reliable and stable with ‌the‌ implementation of ‌firmware updates‌ for our FEB ‌at‌ this stage‌, along with‌ improvements in ‌communication‌ between ‌the‌ DUs and ‌the‌ central station.

\subsection{GP45/65 from October 2024 to date}

\begin{figure*}[]
\centering
\includegraphics[width=0.48\linewidth]{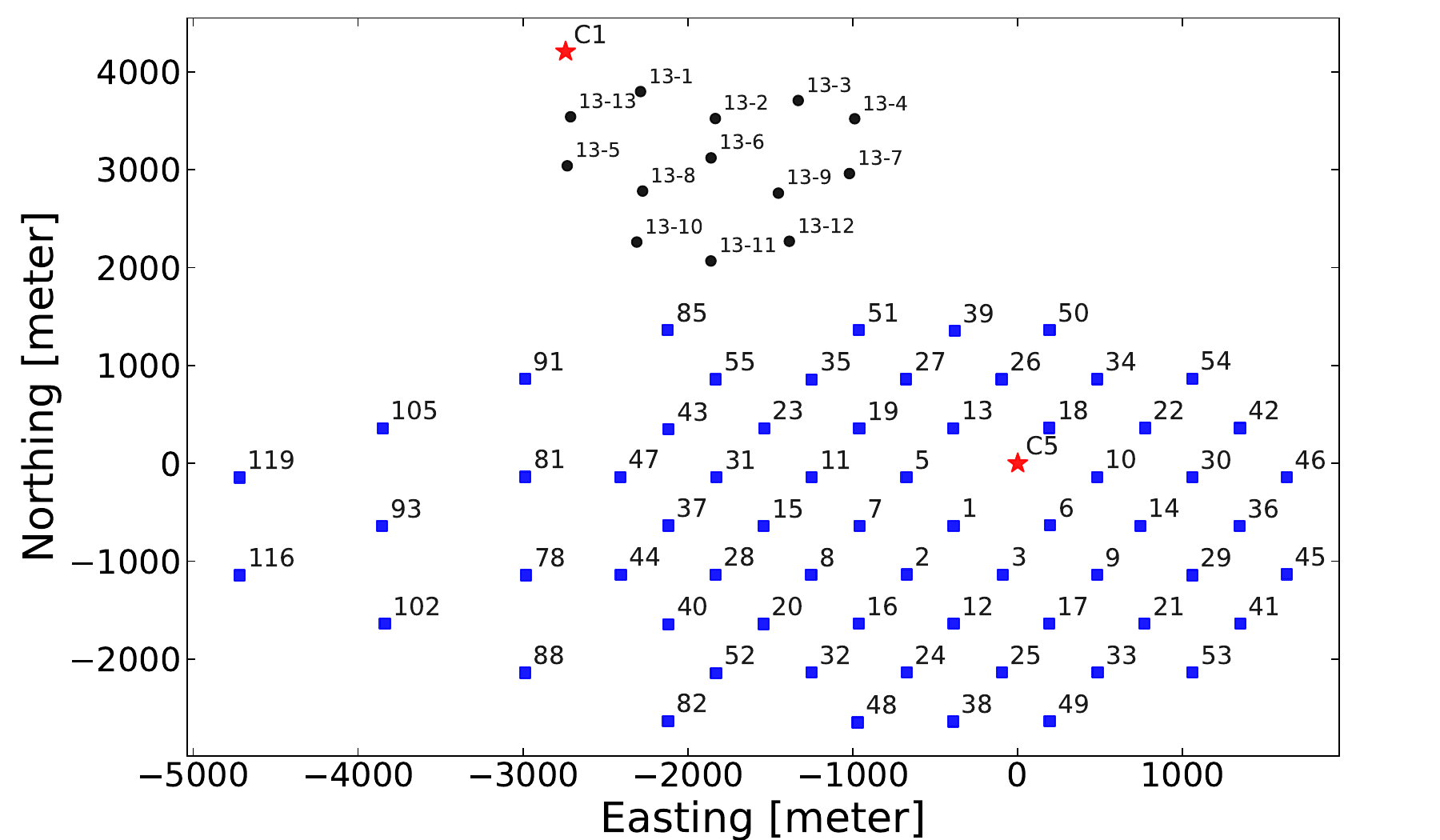}
\includegraphics[width=0.48\linewidth, height=4.7cm]{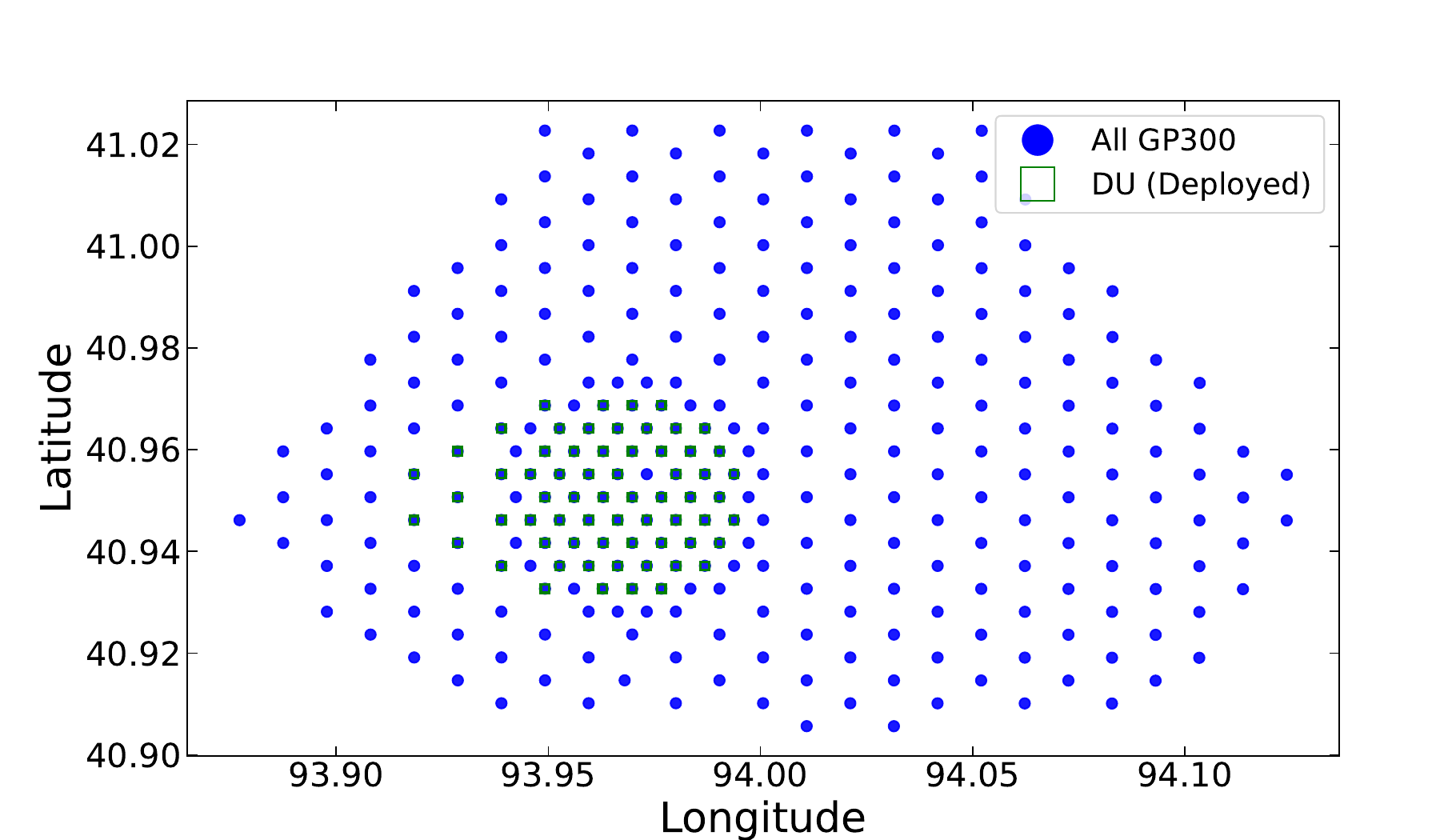}
\caption{Left: In the past two years, black circles and blue squares denote GP13's and GP65's DUs with spacings of ~580 m and ~1000 m, respectively(left panel). Red stars mark current central stations. \mkred{C5/C1 serve as data, logistics, and hardware centers. Right: Positions of all the GP300 DUs indicate in blue circles, the green squares inside denote the deployed GP65.}}
\label{fig:deployment}
\end{figure*}

\mkred{Additional} 65 detection bases and 45 FEBs were deployed in October 2024‌,
this new array ‌performs well in detecting‌ transient signals in ‌the‌ ambient ‌environment‌. We stopped the commissioning run of GP13 and ‌merged‌ it into GP45 in April 2024, deployed‌ more new DUs as well. The latest field trip was conducted June of 2025, all DU of GP65 were completely deployed. A‌ much bigger and complete array ‌has been established. The current layout is shown in Fig.\ref{fig:deployment}, with all blue points representing DUs and two red stars indicating the central stations.


\subsection{Data taking in different period.}


\mkred{The GRAND DAQ system has established a functional data transmission pipeline that meets current operational needs. Deployed at XiaoDuShan site in 2023, it has operated continuously for nearly two years. While the system has received incremental upgrades alongside array expansion, its long-term role as the primary DAQ platform of GP300 remains subject to further performance validation. Stability and efficiency improvements are ongoing.}
The current setup of ‌each DU enables us to collect data in different modes, including monitor/unbiased mode (MD) periodically, triggered mode ‌at‌ single DU level (UD), and coincidence ‌detection‌ among a few DUs (CD) that composes ‌events‌. We gradually ‌achieved‌ stable CD data flow with ‌improved‌ data acquisition ‌capability. Data taking rate of UD ‌can‌ reach around 1.3 kHz. On average, ‌tens of Hz‌ could be ‌handled‌ in the case of CD currently. ‌These data have enhanced‌ our knowledge ‌of‌ the radio background and expectations for detecting cosmic rays at XiaoDuShan in past two years. 

\section{Preliminary results}

\subsection{Preliminary calibration for timing}

The GRAND antenna, called ‌the‌ HorizonAntenna, ‌is‌ optimized for ‌detecting‌ very inclined incident signals.  
Direction reconstruction in GRAND-like experiments relies on the early-late effect across triggered DUs, requiring precise GPS timing. Beacon tests using an LPDA (mounted on a container or mobile pole) revealed non-negligible timing offsets. We focus on a selected test for comprehensive calibration, governed by equation \ref{equ.SNR}, where $Dis_{i}$ and $Dis_{\text{ref}}$ are distances to No. $i$ and the reference DU, respectively, $c$ and $n$ are speed of light and the refractive index of air at site.

\begin{equation}
T_{\text{offset,ns}} = \frac{Dis_{i} - Dis_{\text{ref}}}{c/n} - (T_{i} - T_{\text{ref}})
\label{equ.SNR}
\end{equation}

\begin{figure}[htbp]
  \centering
  \begin{minipage}{0.48\textwidth}
    \centering
    \includegraphics[width=\linewidth]{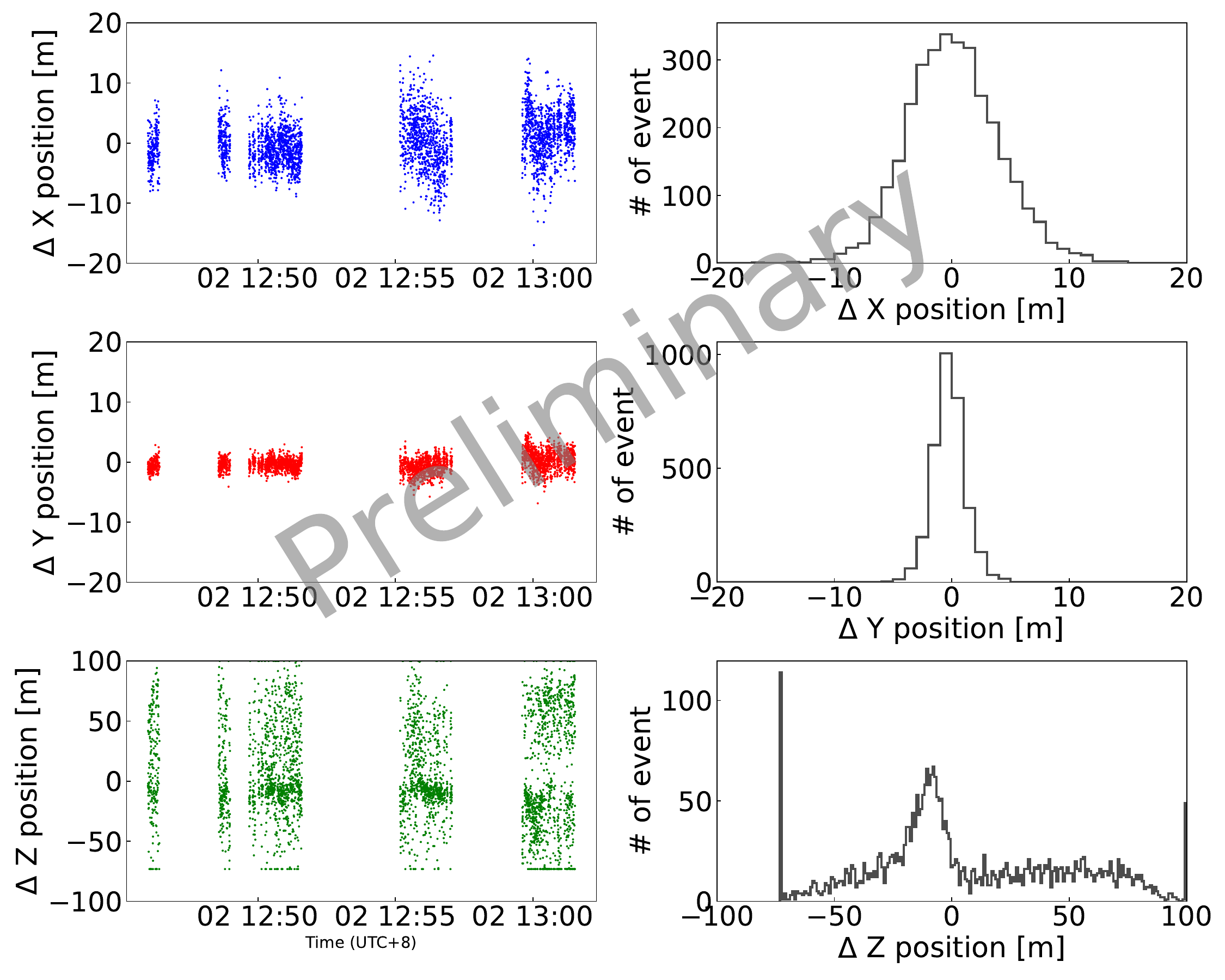}
    \caption{The spatial discrepancies between reconstructed beacon positions and GPS nominal coordinates are analyzed along X/Y/Z axes (top/middle/bottom panels). Temporal and cumulative distributions (left/right panels) demonstrate significantly greater dispersion in the vertical (Z) dimension, seemingly consistent with known GPS altitude measurement uncertainties.}
    \label{fig:timing}
  \end{minipage}
  \hfill
  \begin{minipage}{0.48\textwidth}
    \centering
    \includegraphics[width=\linewidth]{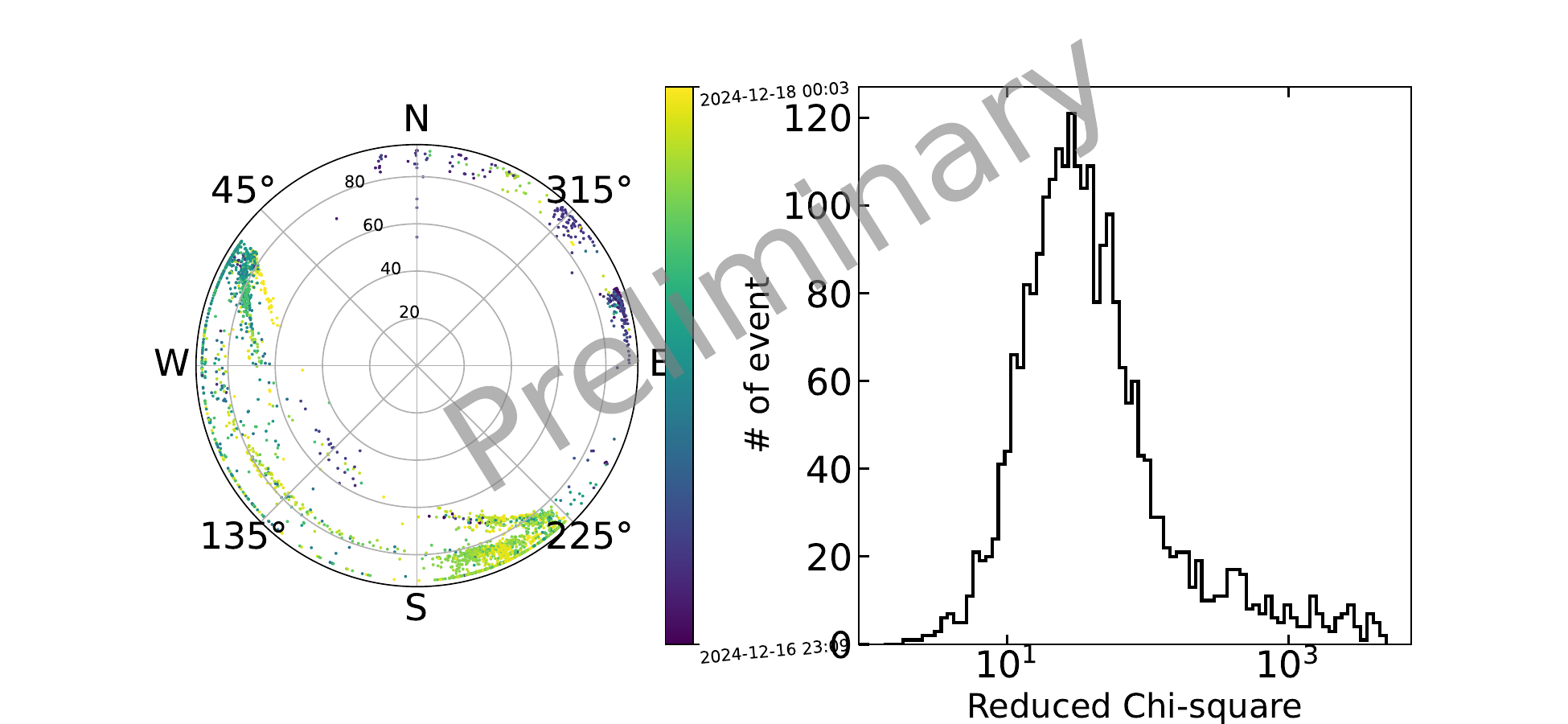}
    \label{fig:timing1}
    \vspace{0.0001cm} 
    \includegraphics[width=\linewidth]{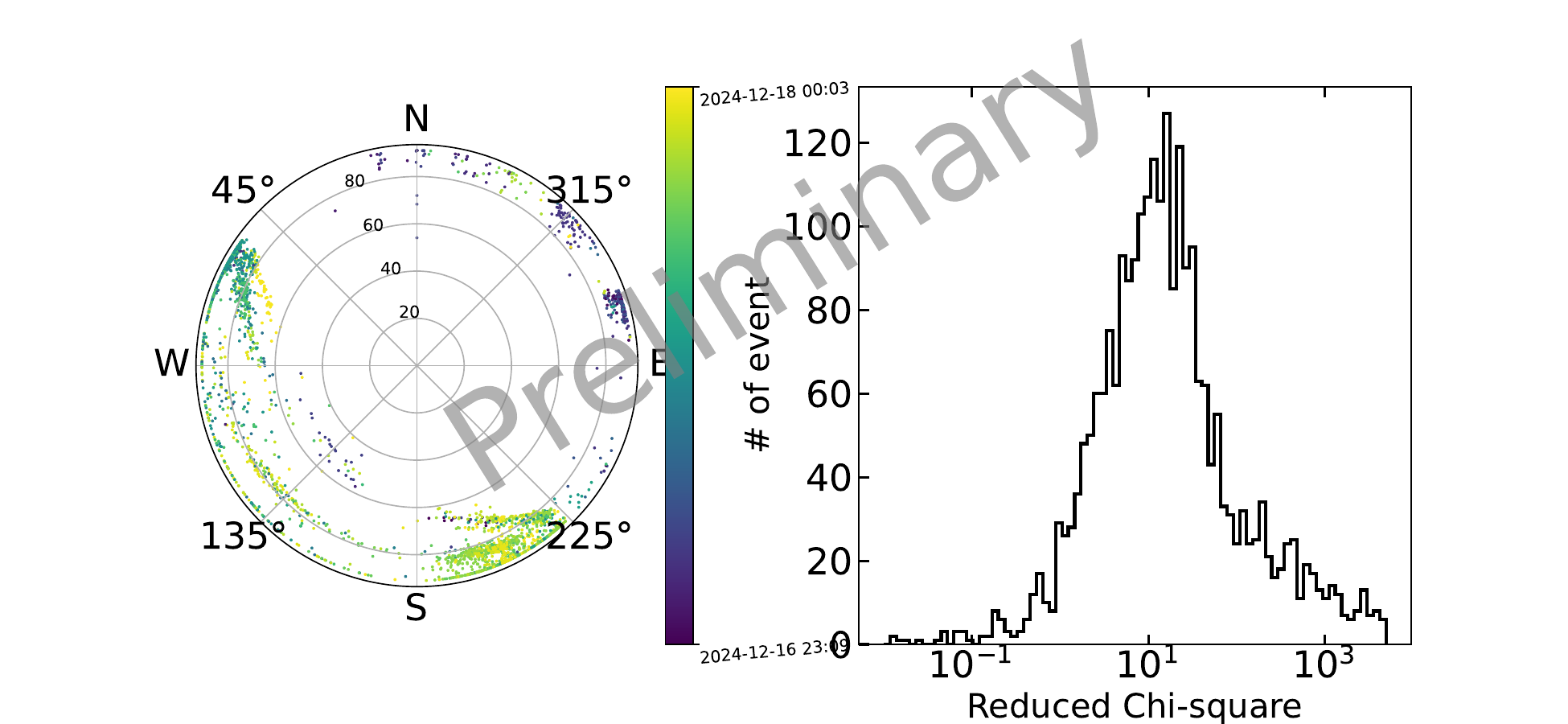}
    \caption{The angular reconstruction performance analysis of December 17, 2024 data shows distinct improvement after offset corrections, with uncorrected results (top) dominated by clustered signals from flights and a nearby transformer station (azimuth ~300°) while corrected results (bottom) demonstrate reduced dispersion, as quantified by the $\chi^2/ndf$ values distributions shown in the right panel.}
    \label{fig:timing2}
  \end{minipage}
\end{figure}

By comparing measured with expected signal arrival times between the beacon emitter and DUs, we determine individual timing offsets $T_{\text{offset,ns}}$. Statistical analysis of multiple test events yields a distribution whose mean and standard deviation represent the systematic GPS timing offset and resolution, respectively. Applying these corrections improves reconstruction performance, as evidenced by small dispersion on ground plane.(Figs. \ref{fig:timing}). Observed timing drifts and glitches \cite{PierreAuger:2015aqe} remain unaddressed in current analysis. Future upgrades will implement a more robust on-site timing system.

We noticed that ‌the‌ quality of angular reconstruction for flight events in Dec 2025 from Fig.\ref{fig:timing2}, ‌conducted‌ one month later than the beacon test‌, still ‌shows‌ better performance when the offset and resolution are considered. \mkblue{This demonstrates that a single execution of the beacon test can still yield valuable insights.}

\subsection{The galaxy contribution at low band on GP300}

\begin{figure*}[]
\centering
\includegraphics[width=7cm]{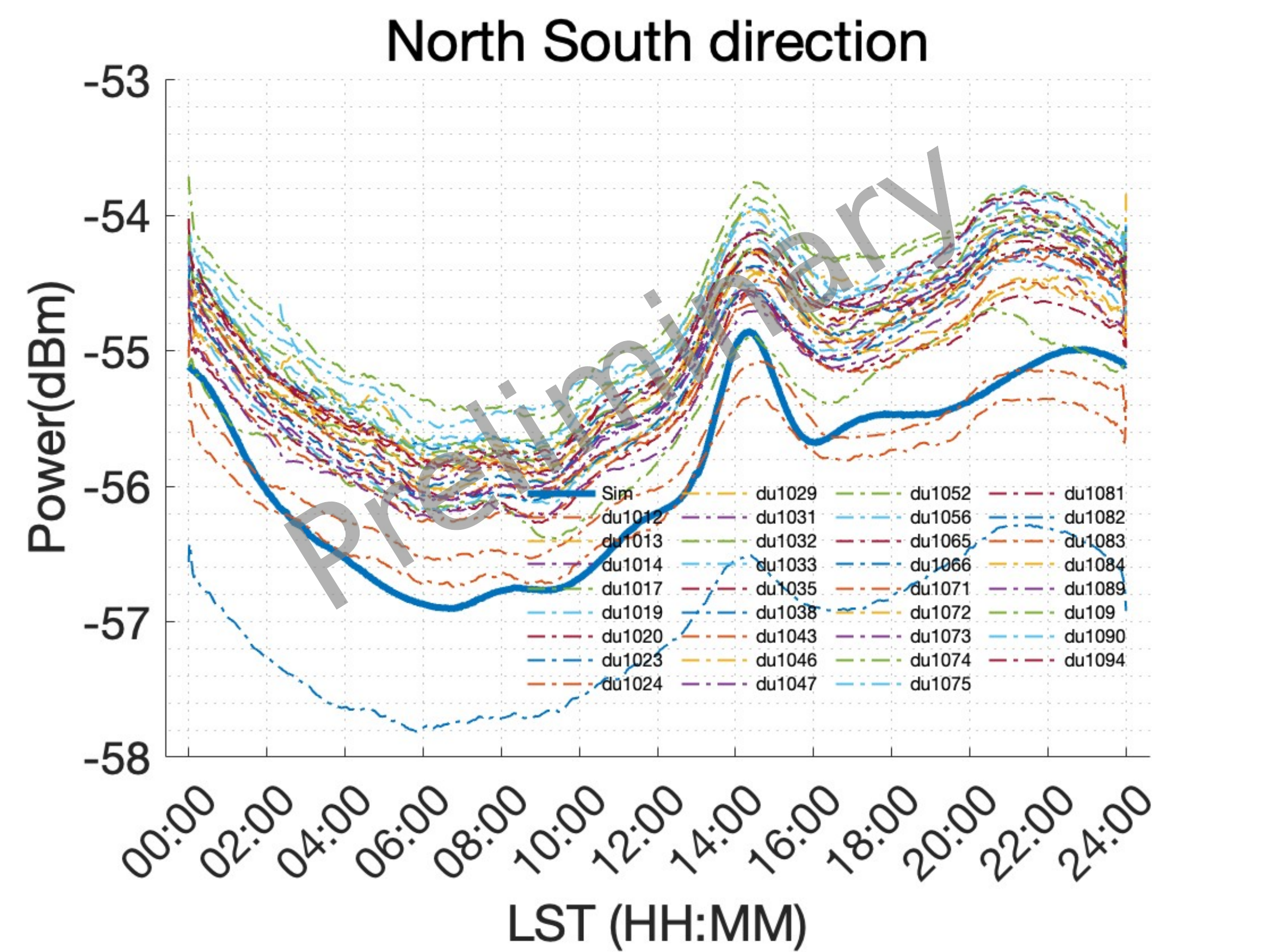}
\includegraphics[width=7cm]{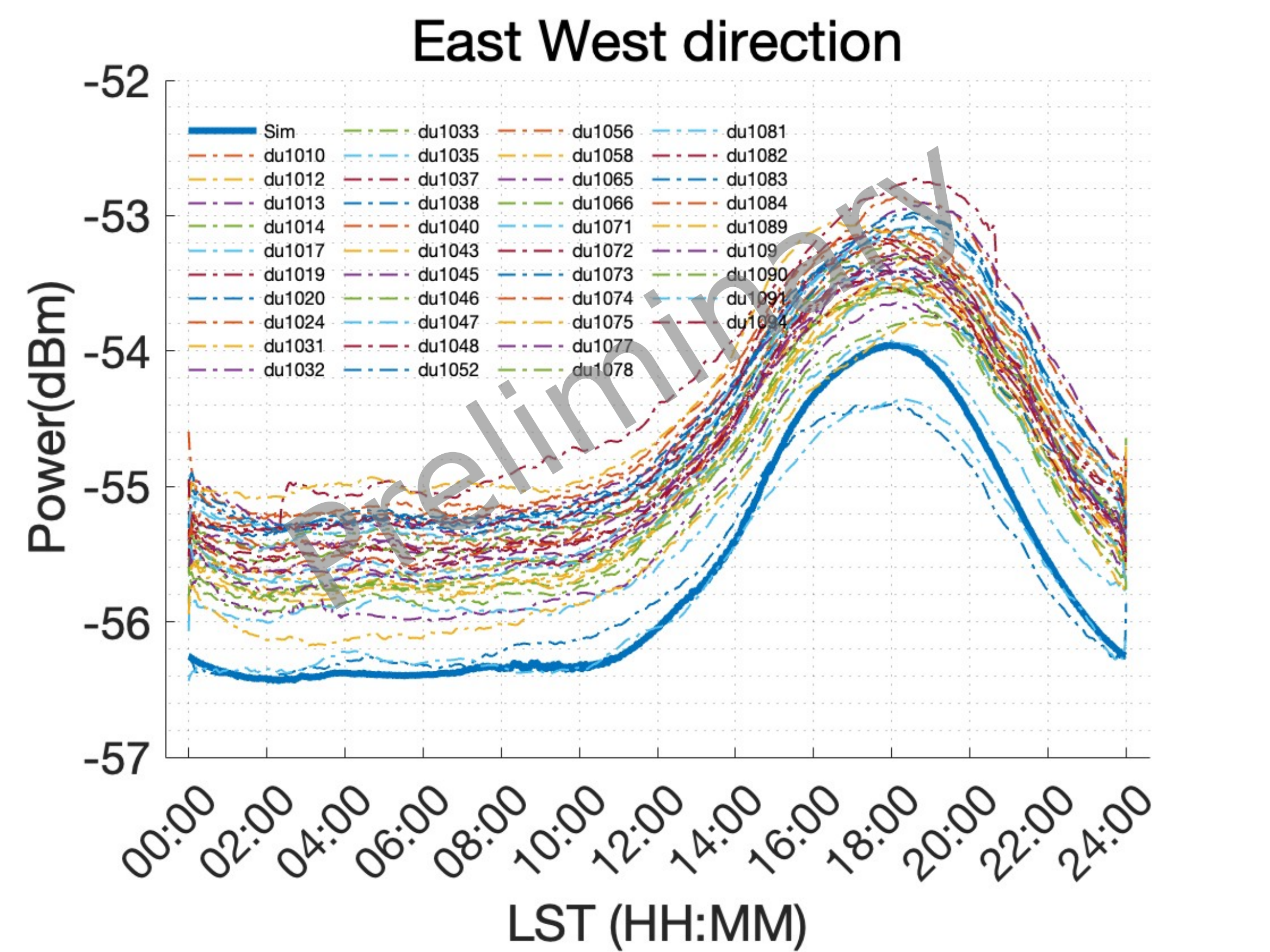}
\caption{The integrated intensity of 60-80 MHz observed by GP45 in April 2025. RF-Chain simulations (thick blue line) show good agreement with experimental data from deployed detector units (dashed lines), displayed separately for N-S and E-W polarizations in left/right panels, respectively.}
\label{fig:galaxy}
\end{figure*}

\mkred{The Galactic radio emission in the GRAND band exhibits time-varying background intensities modulated by the celestial rotation, maintaining consistent Local Sidereal Time (LST) periodicity. Comparative analysis between 60-80 MHz observational data and RF-chain simulations reveals strong modulation alignment (Fig. \ref{fig:galaxy}), with residual discrepancies attributable to gain variations in the FEB's LNA/VGA components.}


The potential rotation angle deviations for each DU from nominal values (N-S/E-W/vertical alignments) critically impact cosmic ray detection accuracy, as they affect both individual antenna radio pattern reconstruction and RF-chain matching. Deployment variations and hardware iterations introduce angular uncertainties requiring systematic calibration - particularly for gain-rotation cross-optimization in our experimental framework.

\subsection{Solar Radio}

\begin{figure}[htbp]
  \centering
  \begin{minipage}{0.48\textwidth}
    \centering
    \includegraphics[width=\linewidth]{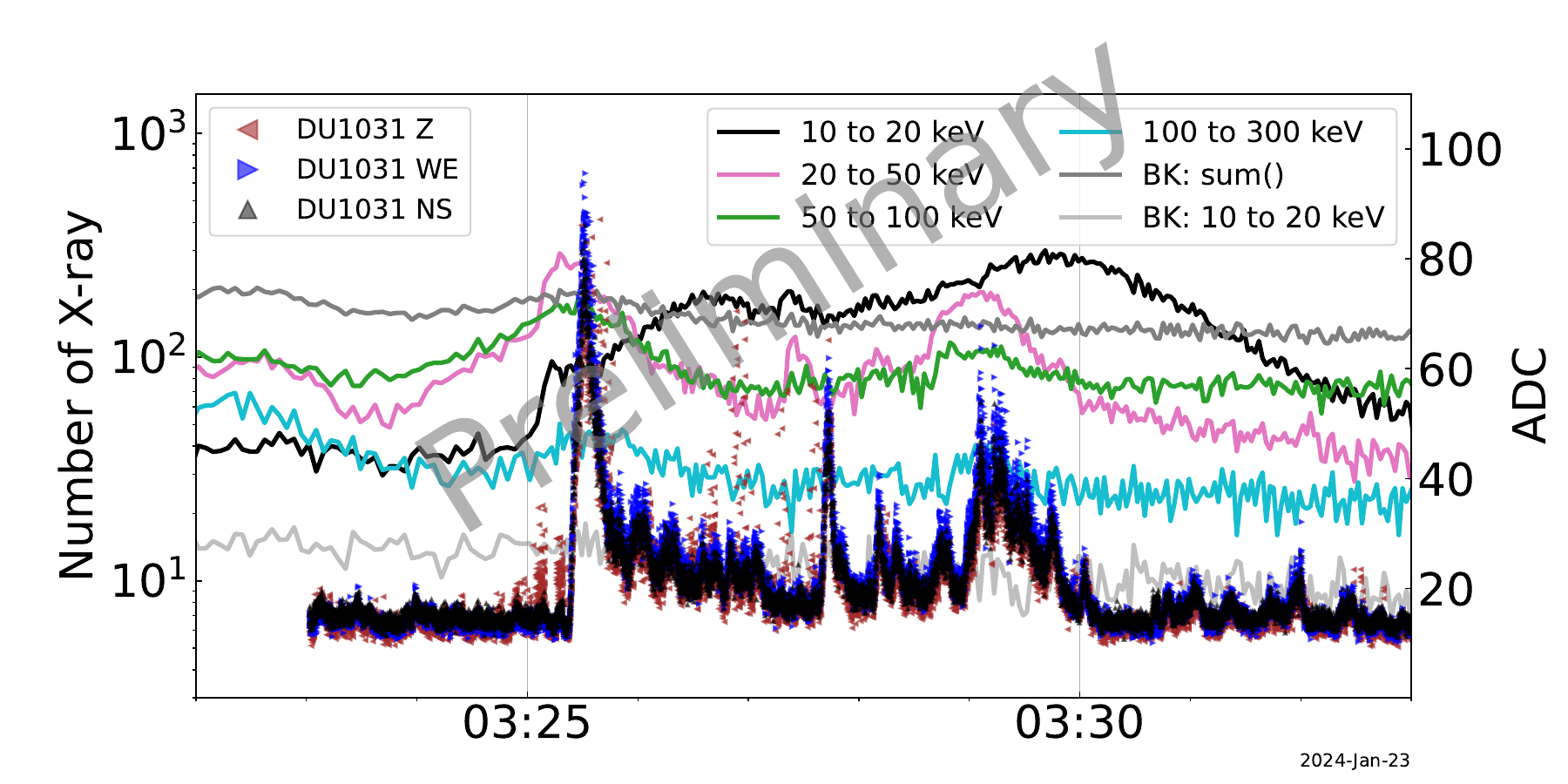}
    \caption{The light curves of solar radio in band of 45 to 105 MHz observed by GP13 in January 2024. The triangle shaped markers present the data from DU1031 at the stage of GP13, the ADC is shown in the right vertical axis. The thin light curves are quick look data of satellite ASO-S in hard X-ray band with vertical axis in left showing the number of density for X-ray photons.}
    \label{fig:solar}
  \end{minipage}
  \hfill
  \begin{minipage}{0.48\textwidth}
    \centering
    \includegraphics[width=8cm]{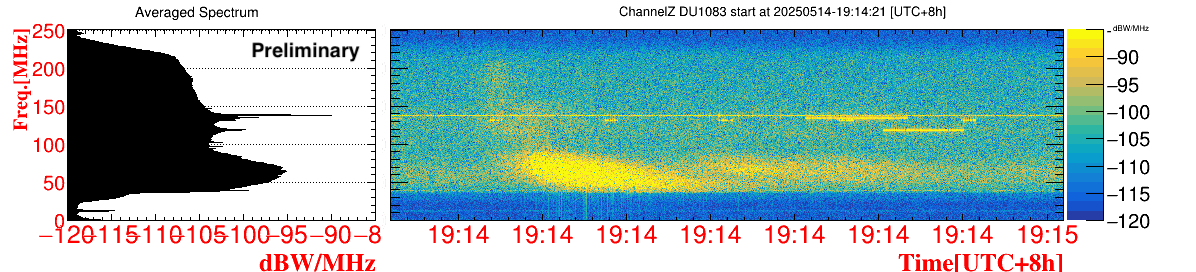}
    \includegraphics[width=8cm]{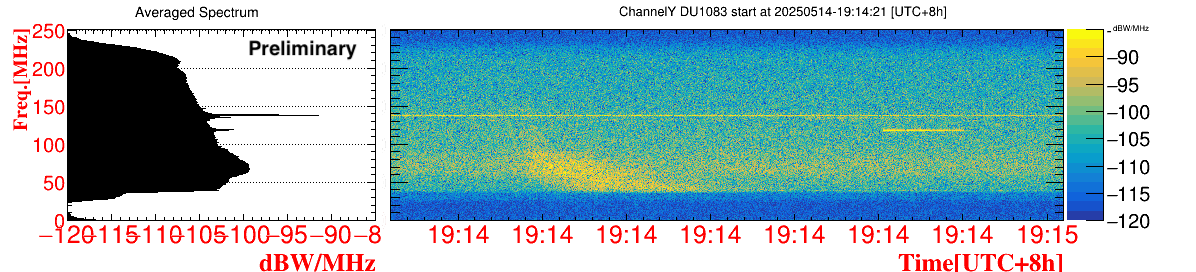}
    \includegraphics[width=8cm]{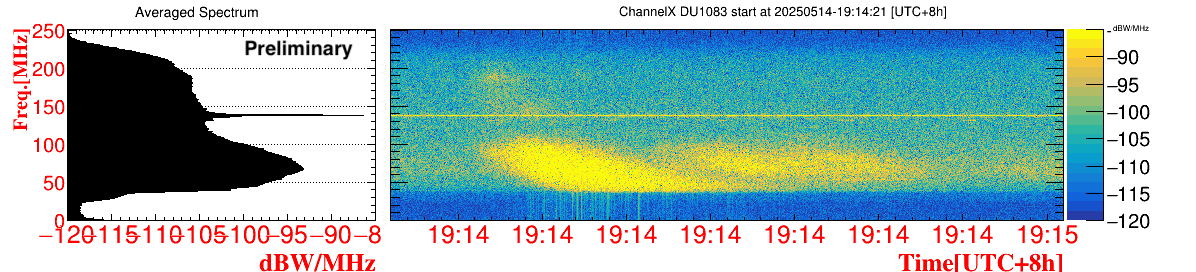}
    \caption{Spectrum from a solar flare at M7.7 class on May 14 2025 with sampling rate of 127 Hz. There are clear excesses below 100 MHz for this event.}
    \label{fig:solar7}
  \end{minipage}
\end{figure}

In the‌ last two years, the Sun ‌has been‌ approaching its latest maximum‌ year of activity, which ‌has emitted‌ many strong ‌radiation events‌, ‌such as‌ solar flares, ‌coronal‌ mass ejections (CMEs), and associated radio radiation ‌across‌ a wide band, from tens of MHz to above GHz. Our antennas ‌have captured‌ solar radio events at different stages. As shown in Fig.\ref{fig:solar}, GRAND-band measurements reveal finer temporal structures and pulsation features compared to hard X-ray counterparts of ASO-S\cite{2022NatAs...6..165G}. Notably, Fig.\ref{fig:solar7} demonstrates tri-polarization spectral characteristics during a M7.7-class flare event.

Solar radio events would provide valuable calibration references for antenna orientation, analogous to galactic emission-based methods. The well-defined planar wavefront characteristics of solar signals enable simultaneous optimization of both antenna gain and radiation pattern parameters during such events.

\subsection{Cosmic rays candidates search}

\begin{figure}[htbp]
  \centering
  \begin{minipage}{0.48\textwidth}
    \centering
    \includegraphics[width=\linewidth]{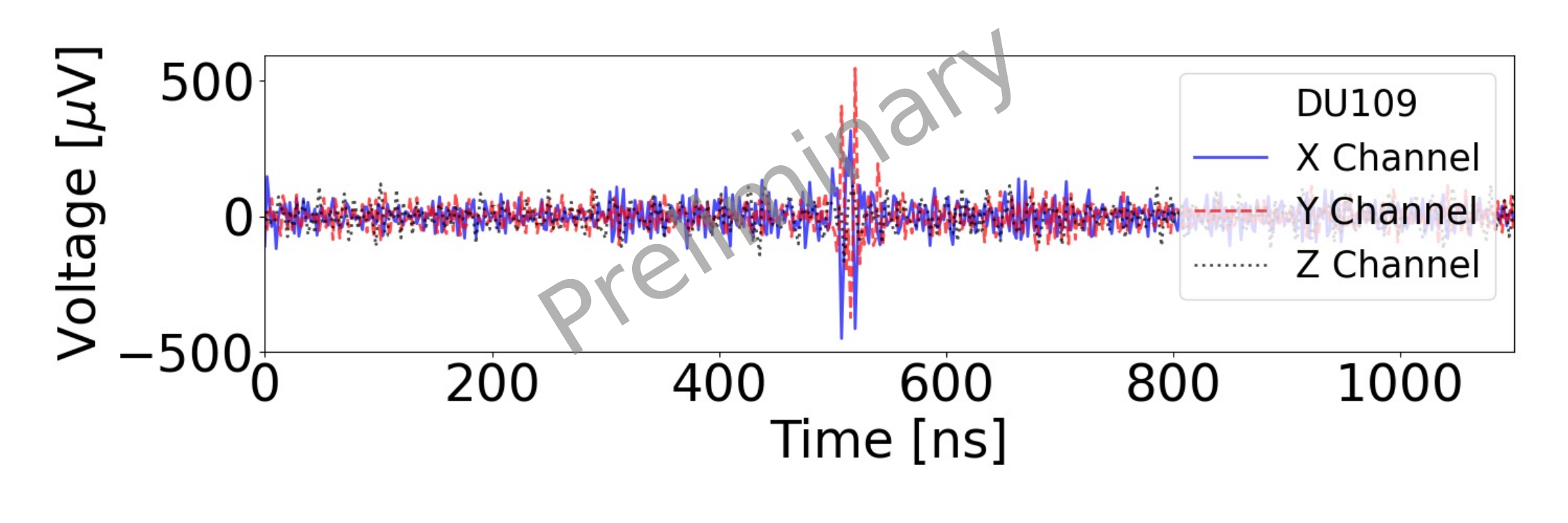}
    \vspace{0mm}
    \includegraphics[width=\linewidth]{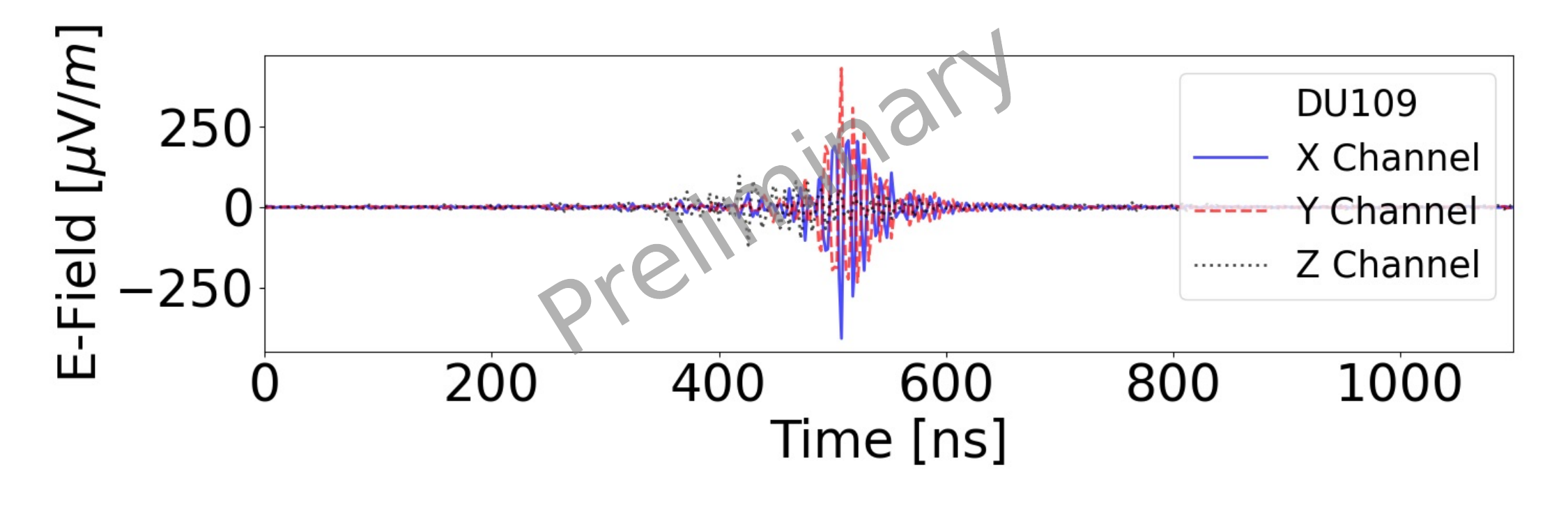}
    \vspace{0mm}
    \caption{Top: One of voltage traces of CR31, detected by GP45. Bottom: Corresponding electronic field trace reconstructed by using the method\cite{Zhang:2025qef}.}
    \label{fig:cr1}
    \end{minipage}
    \hfill
     \begin{minipage}{0.48\textwidth}
    \centering
    \includegraphics[width=\linewidth]{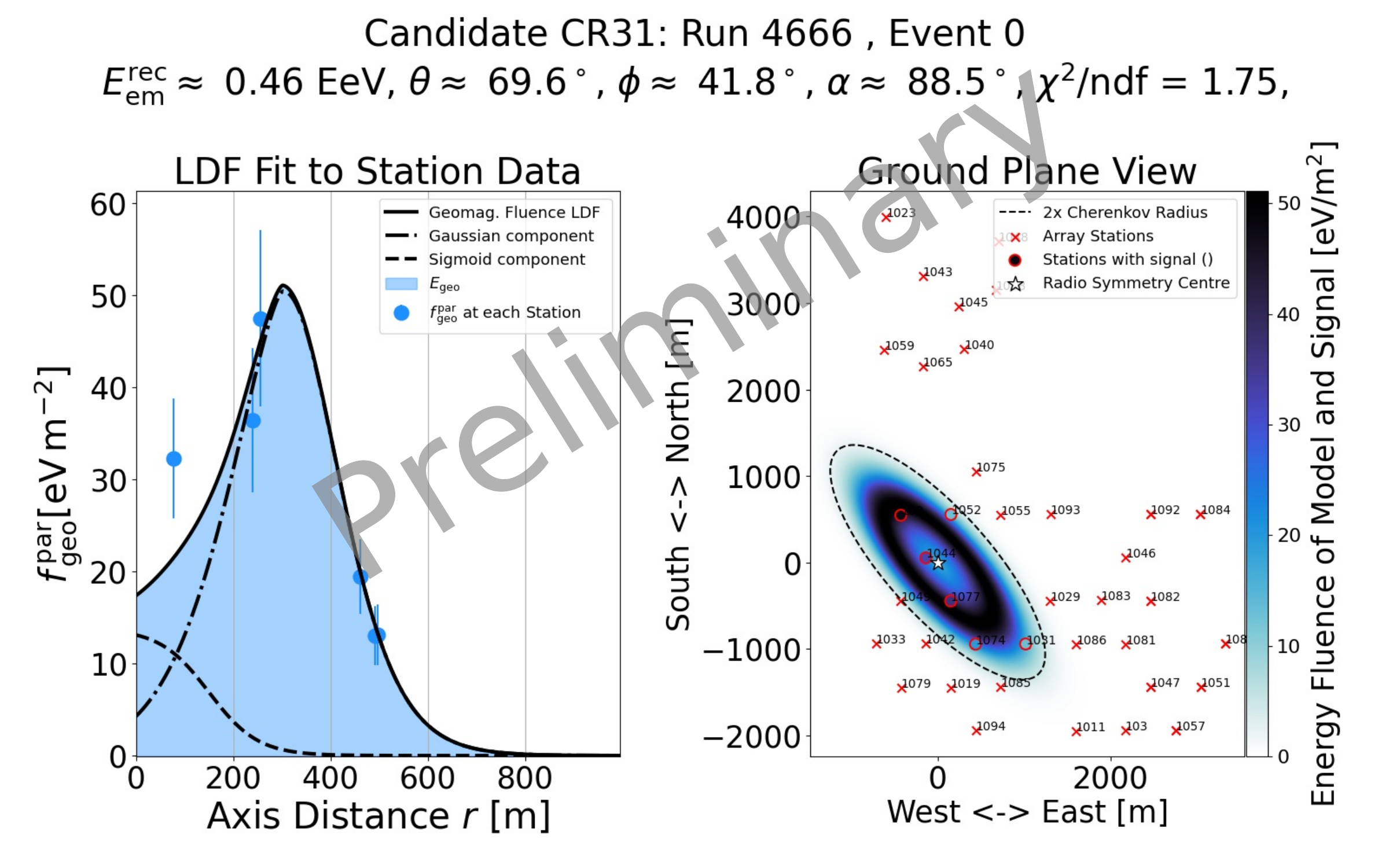}
    \caption{The lateral distribution function fitting (left) and ground plane distribution (right) for CR31\cite{lukas:2025tdh}. The triggered DUs with filled circle indicate the position inside of part of array on the ground plane.}
    \label{fig:cr2}
    \end{minipage}
\end{figure}

\mkred{The GP300 experiment is designed to detect UHECRs with inclined zenith in the energy range of $10^{16.5} - 10^{18}$ $\text{eV}$ using autonomous radio detection techniques. During the past two years, our efforts have focused on hardware upgrades, firmware optimization, and system commissioning. Here we report one of high-confidence cosmic-ray candidate (CR31) identified in the November 2024-March 2025 dataset, with its radio signatures presented in Figs. \ref{fig:cr1} and \ref{fig:cr2}. The candidate selection benefits from: (1) multiple independent reconstruction algorithms, and (2) cross-verification by separate analysis groups. Complete methodology details are available in \cite{Jolan:2025icrc}}

The cosmic-ray detection employs the CD mode as detailed previously. Following GP45's commissioning in November 2024, we initiated online UD operations, recording complete waveform traces at the central station. The trigger algorithm requires:
\begin{enumerate}
\item \textbf{A positive ADC pulse exceeding 5$\sigma$ of the background noise}
\item \textbf{Clean pulse morphology (no pre-/post-pulse artifacts)}
\item \textbf{Coincident triggers from $\geq$5 detector units within a proper window}
\end{enumerate}
During the initial two-month period, online UD ran concurrently with offline CD processing at the central station. Data acquisition rates were limited by three key factors: (1) FEB memory allocation, (2) Array-wide communication bandwidth and (3) DU stability.

A major improvement occurred in February 2025 with the successful deployment of online CD processing, achieving a 10-fold increase in duty cycle ()$\sim$10 Hz) compared to offline operations.

RFI mitigation is crucial for cosmic-ray detection, where we identified significant interference in the 118-140 MHz band and from a northeast transformer station (>150 MHz). We implemented notch filters for the former and are deploying FIR filters (115 MHz above) for the latter. Flight-path RFIs along NW-SE routes show strong weather dependence, constituting most of background events with higher occurrence in overcast conditions, which are most of background events in our data.

\section{Conclusion and Outlook}
The GP300 array has successfully deployed 65 DUs, marking a significant milestone in its phased construction. The system will now enter a sustained data-taking period to validate performance and optimize operation. This stabilization phase is crucial for preparing the next-stage array expansion (targeting 200+ units) and hardware upgrades. Current operations will generate the necessary baseline data for future hardware iterations while maintaining continuous cosmic-ray detection capabilities.
On the other hand, GP300 could enable detection of bright Galactic and nearby fast radio bursts (FRBs). However, the current system can only deliver a small fraction of data points in time domain, which significantly reduces the sensitivity for pulsed signals with high dispersion measures by approximately two to three orders of magnitude. Future upgrades are expected to make FRB detection more feasible for GP300.

\clearpage

\section*{Full Author List: GRAND Collaboration}

\scriptsize
\noindent
J.~Álvarez-Muñiz$^{1}$, R.~Alves Batista$^{2, 3}$, A.~Benoit-Lévy$^{4}$, T.~Bister$^{5, 6}$, M.~Bohacova$^{7}$, M.~Bustamante$^{8}$, W.~Carvalho$^{9}$, Y.~Chen$^{10, 11}$, L.~Cheng$^{12}$, S.~Chiche$^{13}$
% ...
\\
\\
$^{1}$Departamento de Física de Particulas \& Instituto Galego de Física de Altas Enerxías, Universidad de Santiago de Compostela, 15782 Santiago de Compostela, Spain \\
$^{2}$Institut d'Astrophysique de Paris, CNRS  UMR 7095, Sorbonne Université, 98 bis bd Arago 75014, Paris, France \\
$^{3}$Sorbonne Université, Université Paris Diderot, Sorbonne Paris Cité, CNRS, Laboratoire de Physique  Nucléaire et de Hautes Energies (LPNHE), 4 Place Jussieu, F-75252, Paris Cedex 5, France \\
$^{4}$Université Paris-Saclay, CEA, List,  F-91120 Palaiseau, France \\
$^{5}$Institute for Mathematics, Astrophysics and Particle Physics, Radboud Universiteit, Nijmegen, the Netherlands \\
$^{6}$Nikhef, National Institute for Subatomic Physics, Amsterdam, the Netherlands \\
$^{7}$Institute of Physics of the Czech Academy of Sciences, Na Slovance 1999/2, 182 00 Prague 8, Czechia \\
$^{8}$Niels Bohr International Academy, Niels Bohr Institute, University of Copenhagen, 2100 Copenhagen, Denmark \\
$^{9}$Faculty of Physics, University of Warsaw, Pasteura 5, 02-093 Warsaw, Poland \\
$^{10}$Key Laboratory of Dark Matter and Space Astronomy, Purple Mountain Observatory, Chinese Academy of Sciences, 210023 Nanjing, Jiangsu, China \\
$^{11}$School of Astronomy and Space Science, University of Science and Technology of China, 230026 Hefei Anhui, China \\
$^{12}$National Astronomical Observatories, Chinese Academy of Sciences, Beijing 100101, China \\
$^{13}$Inter-University Institute For High Energies (IIHE), Université libre de Bruxelles (ULB), Boulevard du Triomphe 2, 1050 Brussels, Belgium \\
% ...


\subsection*{Acknowledgments}

\noindent
The GRAND Collaboration is grateful to the local government of Dunhuag during site survey and deployment approval, to Tang Yu for his help on-site at the GRANDProto300 site, and to the Pierre Auger Collaboration, in particular, to the staff in Malarg\"ue, for the warm welcome and continuing support.
The GRAND Collaboration acknowledges the support from the following funding agencies and grants.
\textbf{Brazil}: Conselho Nacional de Desenvolvimento Cienti\'ifico e Tecnol\'ogico (CNPq); Funda\c{c}ão de Amparo \`a Pesquisa do Estado de Rio de Janeiro (FAPERJ); Coordena\c{c}ão Aperfei\c{c}oamento de Pessoal de N\'ivel Superior (CAPES).
\textbf{China}: National Natural Science Foundation (grant no.~12273114); NAOC, National SKA Program of China (grant no.~2020SKA0110200); Project for Young Scientists in Basic Research of Chinese Academy of Sciences (no.~YSBR-061); Program for Innovative Talents and Entrepreneurs in Jiangsu, and High-end Foreign Expert Introduction Program in China (no.~G2023061006L); China Scholarship Council (no.~202306010363); and special funding from Purple Mountain Observatory.
\textbf{Denmark}: Villum Fonden (project no.~29388).
\textbf{France}: ``Emergences'' Programme of Sorbonne Universit\'e; France-China Particle Physics Laboratory; Programme National des Hautes Energies of INSU; for IAP---Agence Nationale de la Recherche (``APACHE'' ANR-16-CE31-0001, ``NUTRIG'' ANR-21-CE31-0025, ANR-23-CPJ1-0103-01), CNRS Programme IEA Argentine (``ASTRONU'', 303475), CNRS Programme Blanc MITI (``GRAND'' 2023.1 268448), CNRS Programme AMORCE (``GRAND'' 258540); Fulbright-France Programme; IAP+LPNHE---Programme National des Hautes Energies of CNRS/INSU with INP and IN2P3, co-funded by CEA and CNES; IAP+LPNHE+KIT---NuTRIG project, Agence Nationale de la Recherche (ANR-21-CE31-0025); IAP+VUB: PHC TOURNESOL programme 48705Z. 
\textbf{Germany}: NuTRIG project, Deutsche Forschungsgemeinschaft (DFG, Projektnummer 490843803); Helmholtz—OCPC Postdoc-Program.
\textbf{Poland}: Polish National Agency for Academic Exchange within Polish Returns Program no.~PPN/PPO/2020/1/00024/U/00001,174; National Science Centre Poland for NCN OPUS grant no.~2022/45/B/ST2/0288.
\textbf{USA}: U.S. National Science Foundation under Grant No.~2418730.
Computer simulations were performed using computing resources at the CCIN2P3 Computing Centre (Lyon/Villeurbanne, France), partnership between CNRS/IN2P3 and CEA/DSM/Irfu, and computing resources supported by the Chinese Academy of Sciences.

\end{document}